\documentclass[twocolumn,aps,prl,showpacs, superscriptaddress, amsmath,amssymb]{revtex4}

\usepackage{graphicx}
\usepackage{dcolumn}
\usepackage{bm}
\usepackage{epstopdf}
\usepackage{color}
\usepackage{pdfpages}

\begin{document}

\title{Structures and dynamics of glass-forming colloidal liquids under spherical confinement}
\author{Bo Zhang}
\affiliation{Department of Chemical Engineering and Materials
Science, University of Minnesota, Minneapolis, MN 55455, USA}
\author{Xiang Cheng}
\email{xcheng@umn.edu}
\affiliation{Department of Chemical Engineering and Materials
Science, University of Minnesota, Minneapolis, MN 55455, USA}
\date{\today}
\pacs{64.70.kj, 82.70.Dd, 61.20.Ne} \keywords{colloidal suspensions, glass transtion, liquid structure}

\begin{abstract}

Recent theories predict that when a supercooled liquid approaches the glass transition, particle clusters with a special ``amorphous order'' nucleate within the liquid, which lead to static correlations dictating the dramatic slowdown of liquid relaxation. The prediction, however, has yet to be verified in 3D experiments. Here, we design a colloidal system, where particles are confined inside spherical cavities with an amorphous layer of particles pinned at the boundary. Using this novel system, we capture the amorphous-order particle clusters and demonstrate the development of a static correlation. Moreover, by investigating the dynamics of spherically confined samples, we reveal a profound influence of the static correlation on the relaxation of colloidal liquids. In analogy to glass-forming liquids with randomly pinned particles, we propose a simple relation for the change of the configurational entropy of confined colloidal liquids, which quantitatively explains our experimental findings and illustrates a divergent static length scale during the colloidal glass transition.                

\end{abstract}

\maketitle

Understanding the nature of the glass transition is one of the most challenging problems in condensed matter physics \cite{Berthier11, Wolynes12, Chandler10, Tarjus05}. Although ubiquitous and technically important, glasses still elude a universally accepted theoretical description. A molecular glass forms when the temperature of a liquid is quenched below its glass transition temperature $T_g$. Near the transition, the relaxation of a liquid can slow down by many orders of magnitude with only a modest decrease of temperature by a factor of 2 or 3. The classical thermodynamic theory of Adam and Gibbs suggests that such a super-Arrhenius temperature dependence arises from cooperative particle rearrangements in localized regions that are related to the configurational entropy of supercooled liquids \cite{Gibbs65,Debenedetti01}. 

To illustrate the static correlations associated with these localized regions, ``point-to-set'' correlations have recently been proposed in the framework of the random first-order transition theory (RFOT) \cite{Bouchaud04,Biroli08,Biroli13,Berthier12,Kob12}, which is a modern development of the Adam-Gibbs theory unifying physical insights from the mode coupling theory and the spin glass theory \cite{Berthier11,Wolynes12}. In the RFOT, a gedankenexperiment was conceived \cite{Biroli08}, where particles outside a cavity of radius $R$ in a supercooled liquid are suddenly frozen while the particles inside the cavity are allowed to freely evolve. The point-to-set correlation length, $\xi$, is defined as the minimal $R$ such that the particles at the center of cavity are not affected by the pinning field imposed by the boundary. A cavity with $R<\xi$ constrains the system into a local minimum of the free-energy landscape and captures an ``amorphous order'' particle configuration nucleated within the liquid. Numerical simulations on molecular glass-forming liquids have demonstrated an increase of $\xi$ close to $T_g$ \cite{Bouchaud04,Biroli08,Biroli13,Berthier12,Kob12}. More recently, following a different protocol \cite{Berthier12}, Nagamanasa {\it et al.} have investigated semi-infinite 2D colloidal liquids with a wall of particles pinned in an equilibrated configuration and have experimentally confirmed the existence of static point-to-set correlations away from the pinning wall \cite{Nagamanasa15}. 

Nevertheless, important questions remain unanswered. First, it is still an open question if and how the static correlation develops in 3D experimental systems similar to the gedankenexperiment. Second, recent theory has suggested that, instead of frozen equilibrated boundaries, confined systems with conventional boundaries exhibit a qualitatively similar trend for the development of static correlations \cite{Cammarota13}. Although confined systems have been extensively studied in probing the dynamics of glass-forming liquids \cite{Richert11,Nugent07,Mittal07,Mittal08,Sarangapani08,Sarangapani11,Watanabe11,Emond12,Hunter14}, it has yet to be verified if the predicted static correlation exists in these supposedly familiar systems. Lastly, there still lack systematic experiments that examine how the configurational entropy associated with static correlations varies when a confined colloidal liquid approaches the glass transition. Quantitative understanding of the variation of the configurational entropy is of great importance in shaping the glass transition theory \cite{Berthier11,Gibbs65}. Our experiments aim to address these questions.  

\begin{figure}
\begin{center}
\includegraphics[width=3.35in]{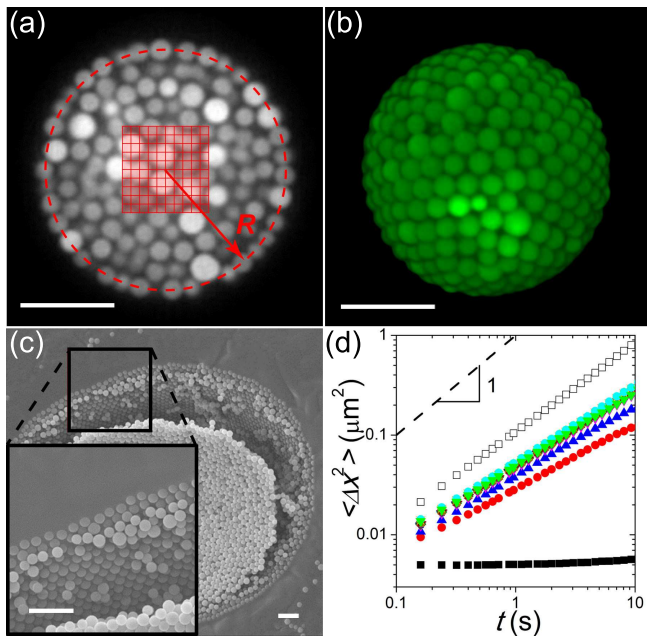}
\end{center}
\caption[Confined colloidal liquids in spherical cavities]{(Color online). Colloidal liquids confined in spherical cavities. (a) A confocal image showing a cross-section of a suspension along the equator of a cavity of radius $R = 5.4d_s$. The 10-by-10 boxes at the center are used for calculating the overlap function, $q_c$. The suspension has $\phi = 0.565$. (b) A 3D reconstruction of the same suspension from a stack of confocal images. (c) A SEM image of a large cavity. The inset shows an enlarged view of the boundary. A layer of particles pinned at the interface can be identified. Scale bars in a-c are 5 $\mu$m. (d) Mean-squared displacements, $\langle\Delta x^2\rangle$, of particles at different distances, $r$, away from the boundary in a cavity of $R = 32.5d_s$. From bottom to top, $r = 0d_s$ (the pinned boundary layer), $1d_s$, $2d_s$, $3d_s$, $4d_s$, $5d_s$, $7d_s$, $9d_s$, $11d_s$, $13d_s$, $15d_s$, $17d_s$ and $19d_s$. The empty squares are for a bulk sample of the same $\phi$. The suspension has $\phi = 0.425$. The boundary layer shows zero motions with the constant indicating the noise level of our tracking algorithm.} \label{Figure1}
\end{figure} 

In our experiments, we use fluorescent polymethylmethacrylate (PMMA) colloidal particles of two different sizes $d_s = 1.29$ $\mu$m and $d_l = 1.64$ $\mu$m (Figs.~\ref{Figure1}a, b), which effectively prevent the crystallization in the system. The polydispersity of each size is smaller than $5\%$ and the number ratio of particles is fixed at 2:1. PMMA particles are suspended in a mixture of decalin and cyclohexyl bromide that matches both the density and refractive index of the particles. To prepare cavities with fixed particle boundaries, we first disperse the colloidal suspension into an aqueous solution of gelatin agent at 70 $^\circ$C to create an oil-in-water emulsion. When temperature is lowered to the room temperature, the aqueous phase solidifies through gelation, which traps a layer of particles at the oil-water interface \cite{supplementary}. The gel has a storage modulus $G' > 10$ MPa, leading to a yield energy, $G'd_s^3$, $10^9$ times larger than the thermal energy $k_BT$. Thus, particles across the interface are permanently pinned, as confirmed by both scanning electron microscopy (SEM) (Fig.~\ref{Figure1}c) and mean-squared displacement (MSD) measurements (Fig.~\ref{Figure1}d). Particle dynamics inside the cavities are recorded using a spinning-disk confocal microscope. The radius of the cavities, $R$, is measured from the average position of the centers of the pinned particles (Fig.~\ref{Figure1}a). We exclude the pinned layer when measuring the volume fraction of the samples, $\phi$ \cite{supplementary}. 

MSDs show a sharp gradient in the slowdown of the particle dynamics near the confining surface. Three particle diameters away from the pinned layer, the particles already exhibit slow yet uniform dynamics (Fig.~\ref{Figure1}d), consistent with the particle dynamics in emulsion droplets with high-viscosity outer fluids \cite{Hunter14}. Thus, the particle dynamics near the center of the cavities reveals the true confinement effect, instead of the wall-induced interfacial effect that is usually present in confined colloidal systems \cite{Nugent07,Mittal07,Mittal08,Sarangapani08,Sarangapani11,Emond12,Watanabe11,Hunter14}.    
  
\begin{figure}
\begin{center}
\includegraphics[width=3.35in]{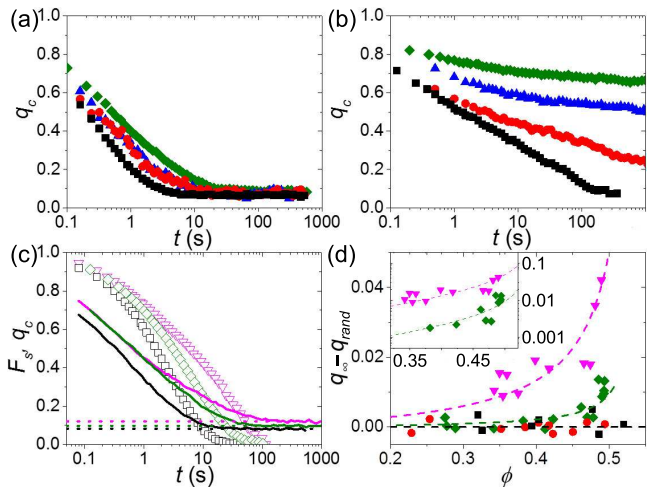}
\end{center}
\caption[Static correlations in 3D confined colloidal liquids]{(Color online). Static correlations in confined colloidal liquids. The time variation of the overlap function, $q_c$, for different cavities with $\phi = 0.41$ (a) and $\phi = 0.54$ (b). From top to bottom, $R = 8.5d_s$ (olive), $16.5d_s$ (blue), $32.5d_s$ (red) and bulk (black). (c) Comparison of self-intermediate scattering functions, $F_s(t)$ (symbols), and $q_c(t)$ (solid lines) at $\phi =0.475$. $R = 4.5d_s$ (magenta), $8.5d_s$ (olive) and bulk (black). $F_s$ reaches 0 when $q_c$ decays to non-zero plateaus (dotted lines). (d) Equilibrium overlap function, $\left(q_\infty-q_{rand}\right)$, versus $\phi$ in the linear scale (main plot) and the semi-log scale (inset). $R = 4.5d_s$ (magenta), $8.5d_s$ (olive), $32.5d_s$ (red) and bulk (black). The dashed lines are visual guides.} \label{Figure2}
\end{figure}

With these specially prepared cavities, we measure the correlation of configurations between the particles at the center of the cavities and the set of particles pinned at the boundary, i.e., the point-to-set correlation. Following the RFOT protocol \cite{Biroli08}, we choose an area of size $3.7d_s \times 3.7d_s$ at the center of the cavities excluding the particles with non-uniform dynamics near the boundary (Fig.~\ref{Figure1}a) \cite{supplementary}. The area is further divided into 10-by-10 small boxes. The box size is small enough such that the number of particles in the box $i$ is either $n_i = 0$ or 1. The overlap function that quantifies the correlation between the initial configuration and the configuration after a time interval $t$ is defined as: 
\begin{equation}
\label{equ1} {q_c(R,t)=\frac{\sum_i\langle n_i(t_0)n_i(t_0+t)\rangle}{\sum_i\langle n_i(t_0) \rangle}},
\end{equation} 
where the average is taken over the initial time $t_0$. Figures~\ref{Figure2}a and b show $q_c(R,t)$ at two different $\phi$. At $\phi = 0.41$, $q_c$ decays to a plateau, which shows a weak dependence on the degree of confinement. However, at $\phi = 0.54$, the confinement exerts a stronger influence on $q_c$. For small cavities with $R \le 16.5d_s$, $q_c$ decays slowly across the entire time window of our experiments.

\begin{figure*}
\begin{center}
\includegraphics[width=7in]{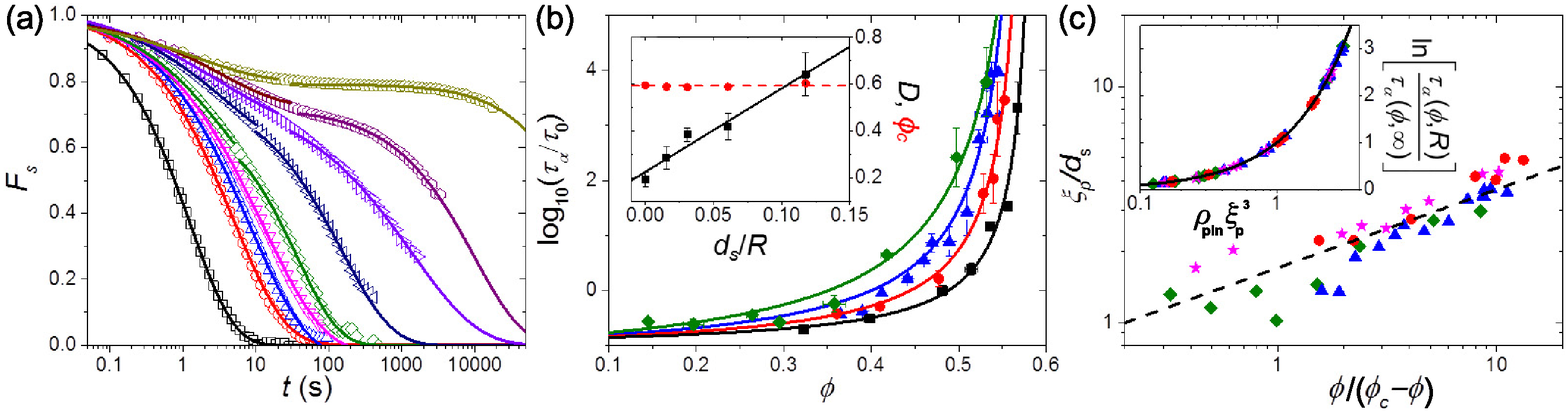}
\end{center}
\caption[Relaxation of confined colloidal liquids]{(Color online). Relaxation of confined colloidal liquids. (a) $F_s(t)$ for $R = 16.5d_s$. From left to right: $\phi =$ 0.390, 0.445, 0.485, 0.505, 0.509, 0.510, 0.515, 0.525 and 0.545. The solid lines are the stretched exponential fittings. Fittings on the highest volume fraction samples serve only as a visual guide. (b) The $\alpha$ relaxation time, $\tau_\alpha$, as a function of $\phi$. From top to bottom, $R = 8.5d_s$, $16.5d_s$, $64.5d_s$ and bulk. $\tau_0 = 2.87$ s is the Brownian relaxation time of small particles in the dilute limit. The solid lines are the VFT fittings. The inset shows the divergent volume fraction $\phi_c$ (red disks) and the fragility index $D$ (black squares) as a function of $d_s/R$. The dashed line indicates $\phi_c = 0.592$ and the solid line is a linear fit. (c) The pinning length $\xi_p/d_s$ vs $\phi/(\phi_c-\phi)$. Besides the same symbols used in (b), the stars indicate $R=32.5d_s$. The dashed line indicates the 1/3 scaling. The inset shows the collapse of $\ln\left[\tau_\alpha(\phi,R)/\tau_\alpha(\phi,\infty)\right]$ with a rescaled variable, $\rho_{pin}\xi_p^3$. The solid line is a linear fit.} \label{Figure3}
\end{figure*}

Two factors determine the time variation of $q_c$: ({\it i}) the development of a true static point-to-set correlation; ({\it ii}) the slowdown of particle dynamics near the glass transition, which deteriorates further due to the effect of confinement \cite{Nugent07,Sarangapani08,Emond12}. To characterize and then exclude the influence of slow particle dynamics on $q_c(t)$, we investigate single particle dynamics through the self-intermediate scattering function: 
\begin{equation}
\label{equ2} {F_s(Q,t)=\langle\frac{1}{N}\sum_i\cos\left(Q\cdot[x_i(t_0+t)-x_i(t_0)]\right)\rangle},
\end{equation}                              
where we choose $Q = 7.4d_s^{-1}$ from the position of the first peak of the structural factor $S(Q)$ \cite{supplementary}, $x_i(t)$ is the location of particle $i$ at time $t$ and $N$ is the number of particles in the studied area. Note that, different from $q_c$, $F_s$ represents a self-correlation, equivalent to the self-overlap function $q_c^{self} \equiv \sum_i\langle n_i^s(t_0)n_i^s(t_0+t)\rangle/\sum_i\langle n_i^s(t_0) \rangle$, where $n_i^s(t_0)n_i^s(t_0+t)=1$ only when the {\it same} particle occupies the cell $i$ at time $t_0$ and $t_0+t$. $F_s$ exhibits a characteristic two-step relaxation above a certain $\phi(R)$ (Fig.~\ref{Figure3}a). We find that $q_c$ reaches the plateau when $F_s$ decays to 0 (Fig.~\ref{Figure2}c) \cite{Berthier12,Ozawa15}. Accordingly, we measure the equilibrium overlap, $q_\infty = q_c(R, t = t^*)$ at a time $t^*$ when $F_s = 0$. The protocol ensures that our measured $q_\infty$ is a true thermodynamic quantity reflecting the structural rather than the dynamic signatures of the system. Specifically, we average the plateaued $q_c$ in a time window over 200 s to reduce statistical errors. Since a bulk sample does not develop the static correlation before the ideal glass transition \cite{Biroli08,Cammarota13}, $q_\infty$ of bulk samples defines $q_{rand}(\phi)$---the overlap between two uncorrelated configurations at a given $\phi$. $q_{rand}(\phi)$ forms the baseline for comparison, which we obtain from both experiments and theoretical and numerical calculations  \cite{supplementary}. 

Fig.~\ref{Figure2}d shows $\left[q_\infty(\phi)-q_{rand}(\phi)\right]$ for different confinements. $q_\infty$ deviates from $q_{rand}$ at $\phi \approx 0.47$ for the cavities of $R = 8.5d_s$, indicating a static point-to-set correlation length $\xi \gtrsim 8.5d_s$, similar to that found in unconfined 2D systems \cite{Nagamanasa15}. Compared with bulk samples of the same $\phi$, particles inside a small cavity equilibrate and sample fewer numbers of states that are compatible with the constraint of pinned boundaries, which, therefore, leads to a larger overlap \cite{Cammarota13}. Different from equilibrated boundaries that pin particles in a single state when $R < \xi$, our confining boundaries allow degenerate states and interstate dynamics \cite{supplementary}. Consistent with the RFOT, the increase of $q_\infty$ with $\phi$ becomes more drastic as $R$ is reduced (Fig.~\ref{Figure2}d). Since the size of the amorphous-order particle clusters increases with $\phi$, a smaller cavity captures the emergence of the clusters at lower $\phi$ and, thus, exhibits a larger $q_\infty$. 

Next, we investigate the influence of the static correlation on the relaxation of confined colloidal liquids. We obtain the relaxation time of the liquids, $\tau_{\alpha}$, by fitting the $\alpha$ relaxation of $F_s$ with stretched exponential functions, $F_s(t) = A\exp[-(t/\tau_\alpha)^\beta]$, where $A$ is the Debye-Waller factor (Fig.~\ref{Figure3}a). $\tau_\alpha$ as a function of $\phi$ for different $R$ is shown in Fig.~\ref{Figure3}b. Although the relaxation is much slower for samples under confinement \cite{Nugent07,Sarangapani08,Emond12}, we find that a fitting of $\tau_\alpha(\phi)$ using the Vogel-Fulcher-Tammann (VFT) relation, $\tau_\alpha(R,\phi) = \tau_0\exp[D\phi/(\phi_c-\phi)]$, gives a constant $\phi_c = 0.59 \pm 0.01$, independent of $R$ (Fig.~\ref{Figure3}b inset). $\phi_c$ is the volume fraction of the apparent divergence of the relaxation time and indicates the ideal glass transition in the Adam-Gibbs theory. Moreover, we find that the fragility index, $D$, increases linearly with $1/R$, following $D(R)=D(\infty)+c(d_s/R)$ with $c = 3.56$ (Fig.~\ref{Figure3}b inset). Bulk samples have the smallest $D$ and exhibit the most fragile behaviors. In addition to the confinement, the change of the fragility of glass-forming liquids has also been reported in experiments varying particle stiffness \cite{Mattsson09} and in simulations varying the curvature of space \cite{Sausset08}, the degree of polydispersity \cite{Kawasaki10} and interparticle potentials \cite{Parmar15}. Although our results are qualitatively similar to previous experiments on 2D confined vibrated granular particles \cite{Watanabe11}, it is worth noting that the change of fragility in our 3D colloidal system is not due to the slowdown of particle dynamics near the wall as proposed in the previous study. When analyzing $q_c$ and $F_s$, we only consider particles with uniform dynamics in the center of the cavities excluding the slow particles near the boundary (Fig.~\ref{Figure1}d).
     
With due caution in interpreting the extrapolated data \cite{Berthier11,Chandler10}, we can gain a quantitative understanding of the configurational entropy of confined colloidal liquids. From the Adam-Gibbs theory, the relaxation of a supercooled liquid is given by $\tau_\alpha \sim \exp\lbrack A_0\phi/s_c \rbrack$, which leads to the VFT relation when the configurational entropy density of the liquid $s_c=K(\phi_c-\phi)$, where $K$ is a numerical constant and the fragility index $D=A_0/K$ \cite{Gibbs65,Sastry01}. $\tau_\alpha$ diverges when $s_c = 0$ and $\phi=\phi_c$. Our experiments show that the configurational entropy of colloidal liquids vanishes at a constant $\phi_c$, independent of confinement. This result agrees with numerical studies on glass-forming liquids with randomly pinned particles \cite{Charkrabarty15}, where the extrapolated Kauzmann temperature $T_K$ is independent of the concentration of pinned particles, $\rho_{pin}$. Note that $\rho_{pin} \approx (d_s/R)(1/d_s^3)$ in our confined system. 

Inspired by this similarity, we propose the following formula for the configurational entropy density of confined colloidal liquids \cite{Charkrabarty15}: $s_c(\phi,R)=F(R)\cdot s_c(\phi,\infty)=F(R)\cdot K(\phi_c-\phi)$, where $s_c(\phi,\infty)$ is the configurational entropy density of bulk samples at $\phi$, and $F(R)\in(0,1]$ is an increasing function of $R$ with $F(\infty) = 1$. The relation is consistent with our experiments, i.e., $s_c(\phi,R) = 0$ at a constant $\phi_c$ independent of $R$. Furthermore, at a finite $R$, the system has a reduced number of states with $s_c(\phi,R) < s_c(\phi,\infty)$, resulting in a non-zero $(q_\infty-q_{rand})$ (Fig.~\ref{Figure2}d). More importantly, under the Adam-Gibbs assumption, the relation leads to $\tau_\alpha(\phi,R) \sim \exp\left( A_0\phi/\left[F(R)K(\phi_c-\phi)\right]\right)$ with the confinement-dependent fragility, $D(R)=A_0/[F(R)K]$, which successfully interprets the decrease of $D(R)$ with increasing $R$ (Fig.~\ref{Figure3}b inset). Quantitatively, we have
\begin{eqnarray}
\ln\left[\frac{\tau_\alpha(\phi,R)}{\tau_\alpha(\phi,\infty)}\right] & = & \lbrack D(R)-D(\infty) \rbrack \left(\frac{\phi}{\phi_c-\phi}\right) \nonumber \\
& = & \left( \left[ D(R)-D(\infty) \right] \frac{R}{d_s} \right) \rho_{pin} \left(\frac{\phi d_s^3}{\phi_c-\phi}\right) \nonumber \\ 
& = & c \rho_{pin} \left(\frac{\phi d_s^3}{\phi_c-\phi}\right). 
\label{eq3}
\end{eqnarray} 
Here, we use the experimental result, $\left[D(R)-D(\infty)\right]=c(d_s/R)$ (Fig.~\ref{Figure3}b inset). Eq.~(\ref{eq3}) can be further written in a scaling form \cite{Charkrabarty15}: 
\begin{equation}
\label{equ4} {\ln\left[\frac{\tau_\alpha(\phi,R)}{\tau_\alpha(\phi,\infty)}\right] = f(\rho_{pin}\xi_p^3)},
\end{equation}   
where $f(x)=x$ and $\xi_p/d_s = \left[c\phi/(\phi_c-\phi)\right]^{1/3}$. $\xi_p$ represents a static length scale---the so-called pinning length---which is related to the point-to-set correlation length $\xi$ through $\xi_p/d_s \sim (\xi/d_s)^{1/3}$ near the random-first-order transition and $\xi_p/d_s \sim \xi/d_s$ away from the transition \cite{Charkrabarty15,Charbonneau12,Cammartoa12}. To directly verify the $\xi_p$ scaling, we manage to collapse $\ln[\tau_\alpha(\phi,R)/\tau_\alpha(\phi,\infty)]$ by using a rescaled variable $\rho_{pin}\xi_p^3$ (Fig.~\ref{Figure3}c inset). $\xi_p(\phi)$ thus obtained indeed follows the predicted 1/3 scaling (Fig.~\ref{Figure3}c). Note that $\xi_p(\phi)$ is an intrinsic property of colloidal liquids, independent of confinement \cite{Charkrabarty15,Charbonneau12,Cammartoa12}. Furthermore, the estimate of $\xi$ based on the $q_c$ measurement gives $\xi \gtrsim 8.5d_s$ at $\phi \approx 0.47$ (Fig.~\ref{Figure2}d), leading to $\xi > \xi_p \approx 2.5d_s$, consistent with the scaling relation between the two length scales.  

Our experiments may help to resolve controversies over static correlations in the glass transition. First, we confirm the numerical and theoretical predictions on the 1/3 scaling of the pinning length \cite{Charkrabarty15,Cammartoa12} and illustrate a divergent static length scale in the colloidal glass transition when $\phi \to \phi_c$. Moreover, we show that glass-forming liquids with randomly pinned particles show quantitatively similar dynamics as colloidal liquids under spherical confinement \cite{Berthier12,Charkrabarty15,Ozawa15,Cammarota13}. Thus, the RFOT can be applied for understanding confined colloidal liquids---an extensively studied subject in colloidal science \cite{Nugent07,Mittal07,Mittal08,Sarangapani08,Sarangapani11,Watanabe11,Emond12,Hunter14}. Our findings contradict the numerical study on hard-sphere particles, where the increase of static correlations is found to be negligible \cite{Charbonneau12}. The results are also different from the theoretical prediction, where $\phi_c$ moves to lower $\phi$ under pinning or confinement \cite{Cammarota13,Cammartoa12}.            

It should be emphasized that although we interpret our results within the context of the RFOT, the experimental findings are independent of specific theoretical descriptions. Therefore, it is necessary to check if our experiments can be explained by other competing theories including dynamical facilitation and geometric frustration models \cite{Chandler10,Tarjus05}. For example, the large curvature of small cavities induces strong geometric frustrations in particle packings \cite{Nelson02,Bausch03}, which modify the particle dynamics that may link to our observations \cite{Tarjus05,Nelson02}. In addition to providing experimental results for assessing general glass-transition theories, our study also provides new insights into the dynamics of confined colloidal liquids and may shed light on the behavior of atomic/molecular liquids under nano-confinements \cite{Richert11}. 

\begin{acknowledgments}
We thank Y. Peng and E. Weeks for discussions and D. Odde, C. Macosko and L. Bai for help with the experiments. The research was partially supported by the ACS Petroleum Research Fund (54168-DNI9) and by the NSF MRSEC Program (DMR-1420013).
\end{acknowledgments}

\end{document}